\documentclass{ws-procs9x6}

\def\simgt{\,\rlap{\lower 3.5pt\hbox{$\mathchar \sim$}}\raise 1pt\hbox {$>$}\,}
\def\simlt{\,\rlap{\lower 3.5pt\hbox{$\mathchar \sim$}}\raise 1pt\hbox {$<$}\,}

\begin{document}

\title{Numerical study of the equation of state for two flavor QCD 
at non-zero baryon density
\footnote{\uppercase{P}resented by \uppercase{S. E}jiri.
\uppercase{T}his work is supported by 
\uppercase{BMBF} grant \uppercase{N}o.06\uppercase{BI}102, 
\uppercase{DFG} grant \uppercase{KA} 1198/6-4, 
\uppercase{PPARC} grant \uppercase{PPA}/a/s/1999/00026
and \uppercase{KBN} grant 2\uppercase{P}03 (06925).}}

\author{S.~Ejiri\rlap,$^{\lowercase{\rm a}}$ 
C.R.~Allton\rlap,$^{\lowercase{\rm b}}$ 
M.~D\"{o}ring\rlap,$^{\lowercase{\rm a}}$ 
S.J.~Hands\rlap,$^{\lowercase{\rm b}}$ 
O.~Kaczmarek\rlap,$^{\lowercase{\rm a}}$ 
F.~Karsch\rlap,$^{\lowercase{\rm a}}$ 
E.~Laermann\rlap,$^{\lowercase{\rm a}}$ \lowercase{and} 
K.~Redlich$^{\lowercase{\rm ac}}$}

\address{$^{\lowercase{\rm a}}$Fakult\"{a}t f\"{u}r Physik, 
Universit\"{a}t Bielefeld, D-33615 Bielefeld, Germany \\
$^{\lowercase{\rm b}}$Department of Physics, University of 
Wales Swansea, Singleton Park, Swansea, SA2 8PP, U.K. \\
$^{\lowercase{\rm c}}$Institute of Theoretical Physics, 
University of Wroclaw,PL-50204 Wroclaw, Poland}

\maketitle

\abstracts{
We discuss the equation of state (EoS) for two flavor QCD at 
non-zero temperature and density. 
Derivatives of $\ln Z$ with respect to quark chemical potential 
$\mu_q$ are calculated up to sixth order. 
From this Taylor series, the pressure, quark number density and associated 
susceptibilities are estimated as functions of temperature and $\mu_q$. 
It is found that the fluctuations in the quark number density 
increase in the vicinity of the phase transition temperature and 
the susceptibilities start to develop a pronounced peak 
as $\mu_q$ is increased. 
This suggests the presence of a critical endpoint in the 
$(\mu_q, T)$ plane.
Moreover, we comment on the hadron resonance gas model, which explains 
well our simulation results below $T_c$.
}

\section{Introduction}
\label{sec:intro}

It is important to study QCD at high temperature and density by 
numerical simulations of lattice QCD. 
In particular, studies of the equation of state (EoS) can provide basic 
input for the analysis of the experimental signatures for quark-gluon 
plasma formation. 
We study the EoS in the low but non-zero baryon number density regime, 
interesting for heavy-ion collisions.
The simulation of QCD at non-zero $\mu_q$ is known to be difficult. 
However, studies based on a Taylor expansion with respect to $\mu_q$ 
turned out to be an efficient technique to investigate the 
low density regime \cite{us02,us03}. 

Thermal fluctuations near the critical temperature $T_c$ are 
important information for the understanding of the QCD phase diagram. 
Baryon number fluctuations are expected to diverge at 
the endpoint of the first order chiral phase transition line. 
The electric charge fluctuation in heavy-ion collisions is 
one of the most promising experimental observables to identify 
the critical endpoint.
The fluctuations of quark number, isospin number 
and electric charge are estimated by the corresponding 
susceptibilities, $\chi_q, \chi_I$ and $\chi_C$, which are given by 
$\chi_q= \partial^2 p / \partial \mu_q^2$ and  
$\chi_I= \partial^2 p / \partial \mu_I^2$, 
where $\mu_q=(\mu_u+\mu_d)/2$, $\mu_I=(\mu_u-\mu_d)/2$, and 
$\mu_{u[d]}$ is the chemical potential for the $u[d]$ quark.
For $\mu_u=\mu_d$, the charge susceptibility is 
$\chi_{C} = \chi_q/36 + \chi_I/4$. 

In this study, we investigate the pressure $p$, 
quark number susceptibility $\chi_q$ and isovector 
susceptibility $\chi_I$ by calculating the Taylor expansion 
coefficients of $p$, $c_n$ and $c_n^I$, up to sixth order in $\mu$, 
which are defined by 
$p/T^4 =(\ln Z)/(VT^3) 
\equiv \sum_{n=0}^{\infty} c_n (\mu_q/T)^n$ and 
$\chi_q [\chi_I]/T^2 \equiv 
\sum_{n=2}^{\infty} n(n-1) c_n [c_n^I] (\mu_q/T)^{n-2}$ 
for $\mu_u=\mu_d$.
where $c_n = (\partial^n \ln Z / \partial \mu_q^n)/n!$ and
$c_n^I = (\partial^n \ln Z / \partial \mu_q^{n-2} \partial \mu_I^2)/n!$. 
Moreover, we compare these results with the prediction from
the hadron resonance gas model \cite{KRT1,KRT2}.

\section{Quark gluon gas or hadron resonance gas}

We expect the equation of state approaches that of a free quark-gluon 
gas (Stefan-Boltzmann (SB) gas) in the high temperature limit. 
The coefficients in the SB limit 
for $N_{\rm f}=2$, $\mu_I=0$ 
are well known as 
$c_2=c_2^I=1, c_4=c_4^I =1/(2\pi^2)$ and $c_n=c_n^I=0$ for $n \geq 6$. 

On the other hand, in the low temperature phase QCD is well modelled 
by  a hadron resonance gas. 
If the interaction between these hadrons can be neglected, 
the pressure is obtained by summing over the contributions from 
all resonance states of hadrons.
The contribution to $p/T^4$ from a particle which has mass $m_i$, 
baryon number $B_i$ and third component of isospin number $I_{3i}$ is given by 
\begin{eqnarray}
\frac{p_{m_i}}{T^4}=\frac{1}{2 \pi^2} \left( \frac{m_i}{T} \right)^2 
\sum_{l=1}^{\infty} \frac{\eta^{l+1}}{l^{2}} 
K_2 \left( \frac{lm_i}{T} \right) 
\exp \left[ l \frac{3B_i \mu_q + 2I_{3i} \mu_I}{T} \right] , 
\label{eq:frg}
\end{eqnarray}
where $K_2$ is the modified Bessel function, 
and $\eta$ is $1$ for mesons and $-1$ for baryons.
Moreover, since $m_i/T \gg 1$ for all baryons and 
$K_2(x) \approx \sqrt{\pi/2x} \exp(-x)$, 
the first term of $l$ is dominant in the baryon sector.
Therefore, the pressure can be written by 
$p/T^4=G(T)+F(T) \cosh (3 \mu_q/T)$
for $\mu_I=0$, where $G(T)$ and $F(T)$ are the mesonic and baryonic 
components of $p/T^4$ at $\mu_q=0$, respectively \cite{KRT2}. 
Similarly, we obtain $\chi_q/T^2=9F(T) \cosh (3 \mu_q/T)$ and 
$\chi_I/T^2=G^I(T)+F^I(T) \cosh (3 \mu_q/T)$. 
Here, the mesonic component for $\chi_q$ is zero because mesons 
$(B_i=0)$ are independent of $\mu_q$. 
Therefore, $c_4/c_2=3/4$, $c_6/c_4=c_6^I/c_4^I=3/10$ in the 
region where the non-interacting hadron 
resonance gas provides a good approximation. 

\begin{figure}[t]
\centerline{\epsfysize=1.55in\epsfbox{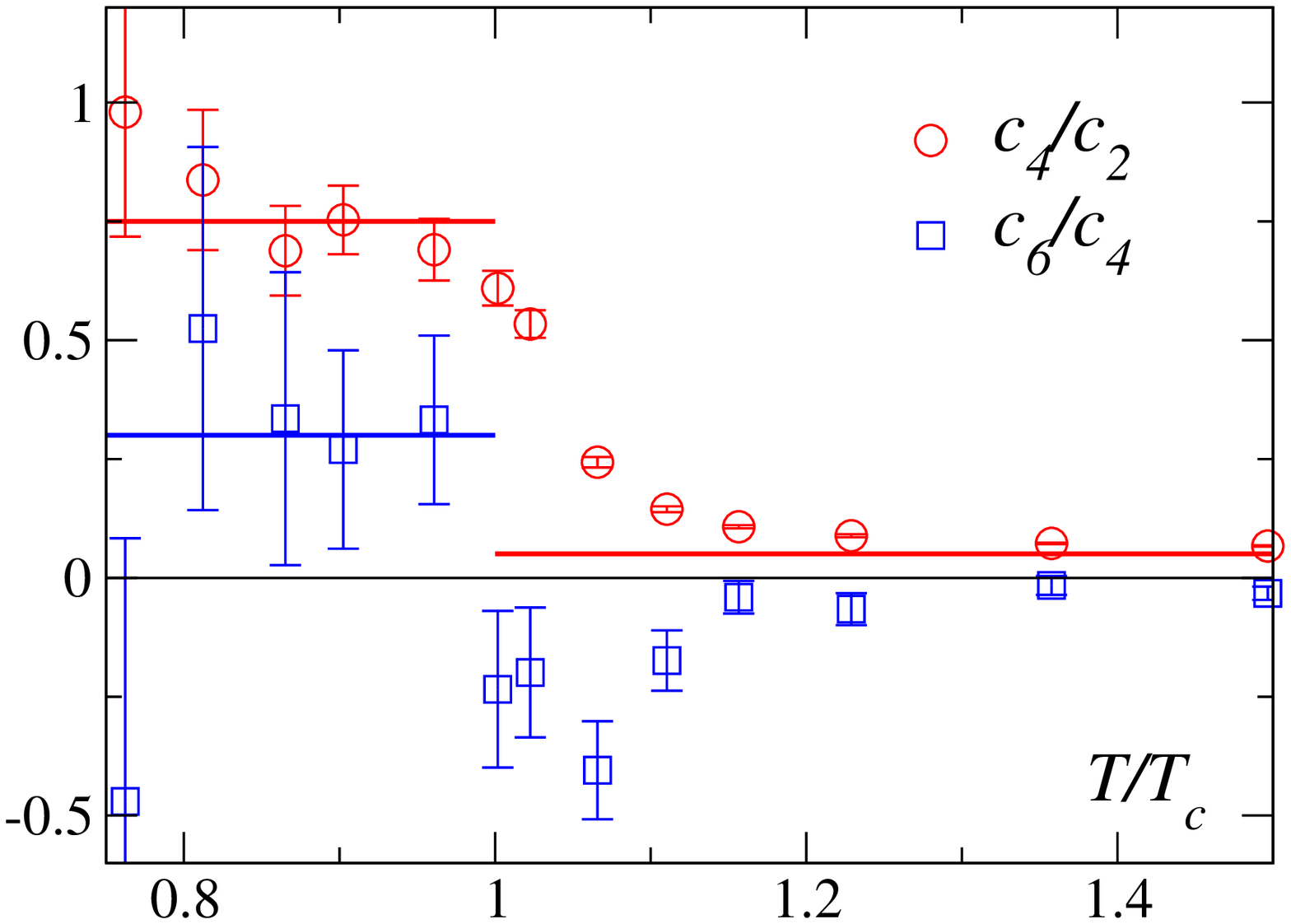}
            \epsfysize=1.55in\epsfbox{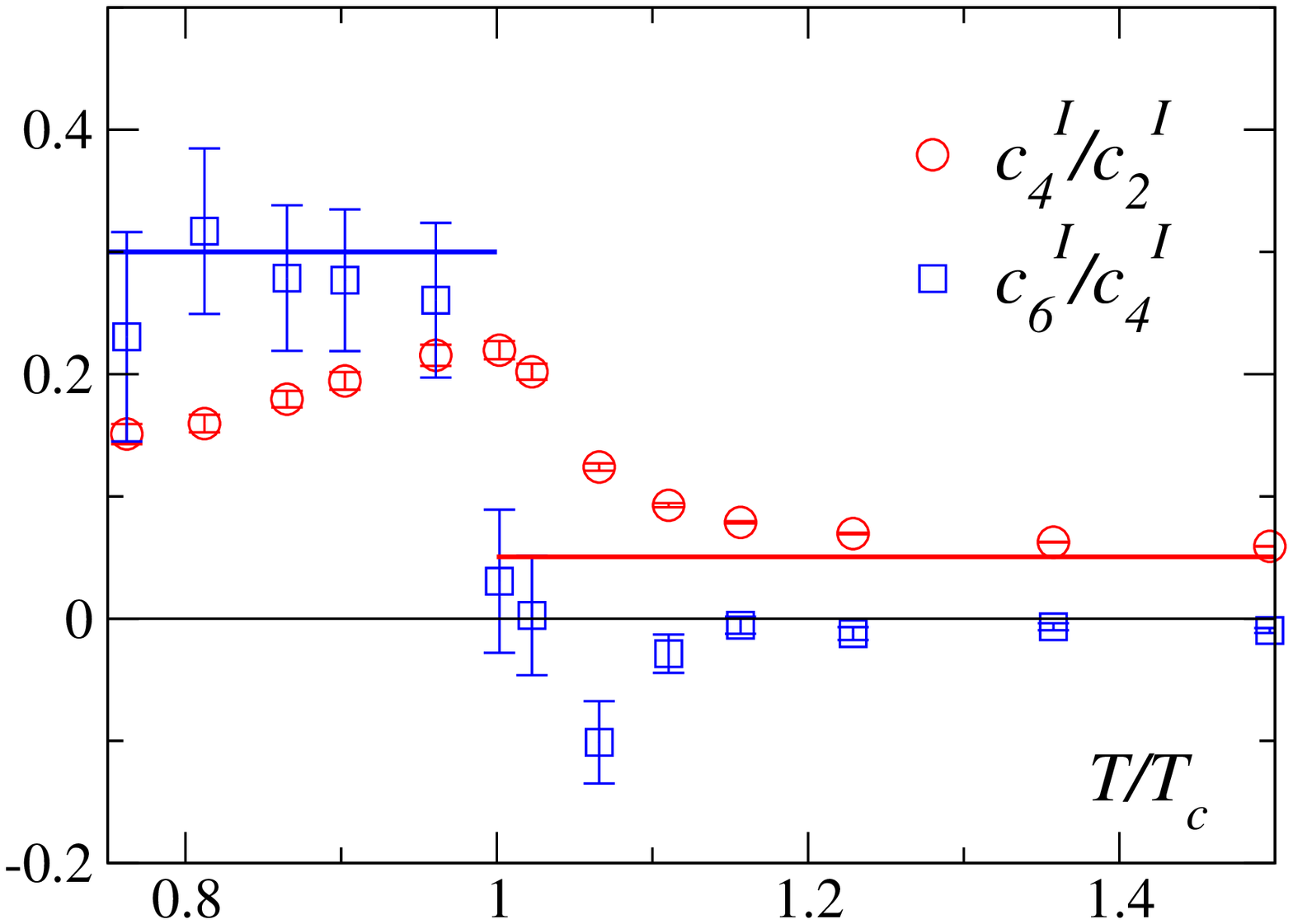}}   
\caption{The ratio of Taylor expansion coefficients for $\chi_q$ 
(left) and $\chi_I$ (right).
\label{fig:c2c4c6}}
\end{figure}

We investigate these coefficients. 
Simulations are performed on a $16^3 \times 4$ lattice for 
2 flavor QCD. The $p4$ improved action \cite{HKS} is employed at 
$ma=0.1$, which gives $m_{PS}/m_{V} = 0.7$. 
The number of configurations is 1000-5000 for each $T$.
We use the random noise method \cite{us02,us03}. 
The results for $c_{n+2}/c_n$ and $c_{n+2}^I/c_n^I$ 
are shown in Fig.~\ref{fig:c2c4c6}. 
We find these results are consistent with the prediction from 
the hadron resonance gas model for $T/T_c \leq 0.96$ and
approach the SB values, 
i.e. $c_4/c_2= c_4^I/c_2^I=1/2\pi^2$, $c_6/c_4=c_6^I/c_4^I=0$, 
in the high temperature limit. 
These results suggest that the models of free quark-gluon gas
and hadron gas seem to explain the behavior of thermodynamical 
quantities well except in the narrow regime near $T_c$.

\begin{figure}[t]
\centerline{\epsfysize=1.6in\epsfbox{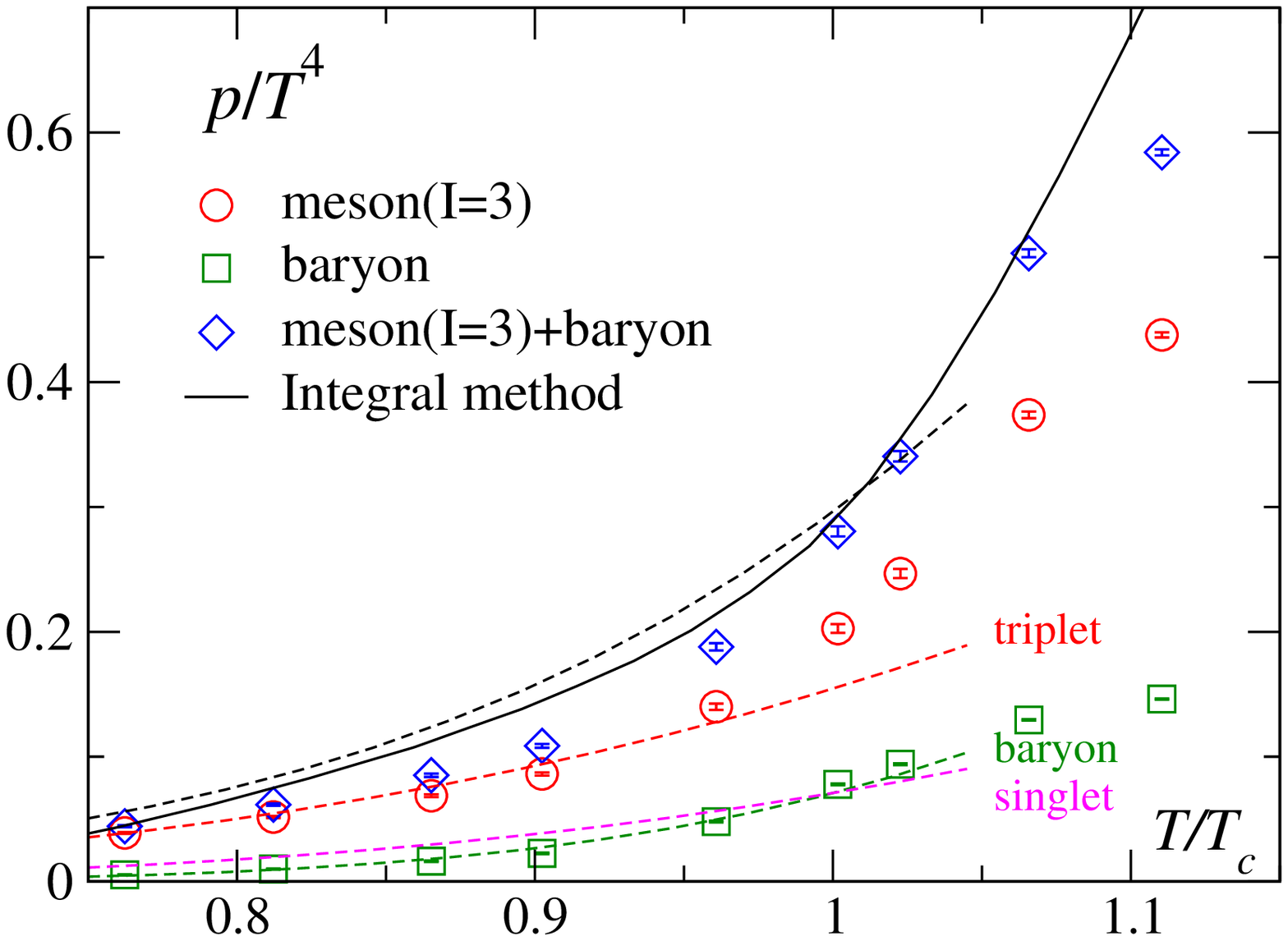}
            \epsfysize=1.6in\epsfbox{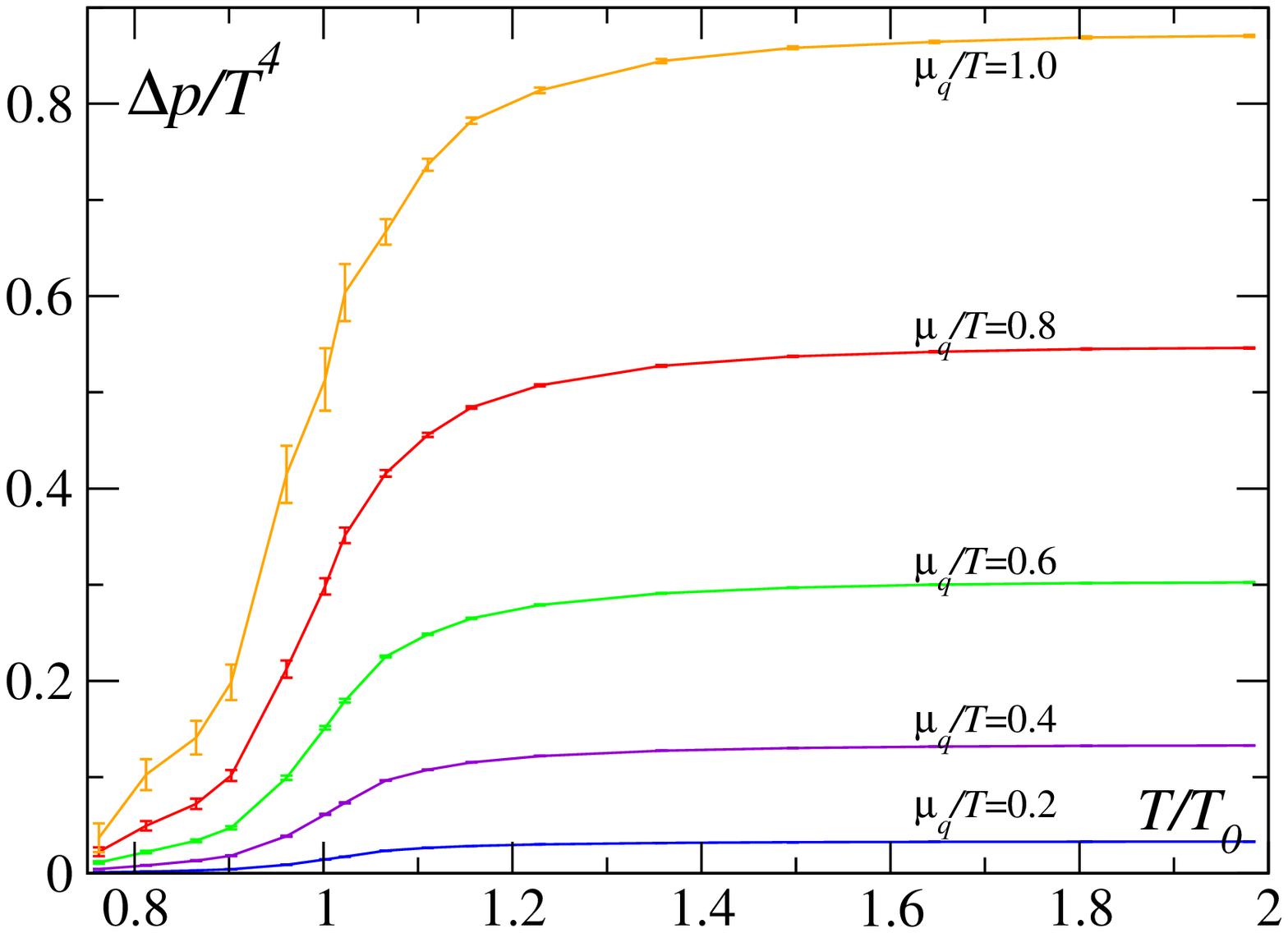}}   
\caption{Pressure at $\mu_q=0$ as a function of $T$ (left), 
and difference of pressure at non-zero $\mu_q/T$ from $\mu_q=0$ 
for each fixed $\mu_q/T$ (right). 
$T_0$ is $T_c$ at $\mu_q=0$.
\label{fig:pres}}
\end{figure}

Moreover, if the data is consistent with the hadron resonance gas model, 
we can separate $\chi_I$ into the contributions from isotriplet mesons 
and baryons, since $G^I+F^I=2c_2^I$ and $9F^I/2=12c_4^I$. 
Isosinglet mesons do not contribute to $\chi_I$ because $I_{3i}=0$. 
Also, $9F=2c_2$ from the equation of $\chi_q$.
Furthermore, for the case of $m_{\pi} \gg T$, the first term $(l=1)$ 
in Eq.(\ref{eq:frg}) is dominant. 
This approximation holds for our present simulations where 
$m_{\pi}/T \simeq 4$-$5$ but will no longer be valid with physical 
quark mass values. For our present simulation parameters, 
the ratio of 
$\partial^2 (p_{mi}/T^4)/\partial (\mu_I/T)^2$ and $p_{mi}/T^4$ equals 
$(2I_{3i})^2$ for $I_{3i}=\{-1,0,1\}$. 
We thus find for the contribution of the triplet meson 
part: $\tilde{G}=3G^I/8$.
We then can estimate the contribution of the 
isosinglet mesons to $p/T^4$ by comparing 
$F+\tilde{G}$ with the total pressure. 
In Fig.~\ref{fig:pres} (left), we plot the baryon component 
$F=2c_2/9$ (square), the triplet meson component 
$\tilde{G}=3c_2^I/4-c_4^I$ (circle) and 
the total (diamond). The solid line is the pressure obtained with 
the integral method in Ref.~\refcite{KLP}. 
The dashed lines in Fig.~\ref{fig:pres} (left) represent the predictions from 
the hadron resonance gas model for the mass parameter of our simulations. 
The resonance states for this quark mass are adjusted by the method 
described in Ref.~\refcite{KRT1}.
The simulation results for $p/T^4$ obtained with the integral method, 
together with the triplet meson and baryon contributions, are 
surprisingly well reproduced by this model calculation. 
The result of the singlet meson contribution may explain the 
difference in the pressure obtained from the integral method and 
that from the Taylor expansion which neglects the singlet part.

\section{Pressure and susceptibilities at $\mu_q \neq 0$} 
\label{sec:sus}

Next, we calculate pressure and quark number susceptibility in a range 
of $0 \leq \mu_q/T \leq 1$, using the data of $c_n$ and $c_n^I$; 
$\Delta(p/T^4) \equiv p(T,\mu_q)/T^4-p(T,0)/T^4=
c_2(\mu_q/T)^2+c_4(\mu_q/T)^4+c_6(\mu_q/T)^6$, 
$\chi_q/T^2=2c_2+12c_4(\mu_q/T)^2+30c_6(\mu_q/T)^4$, 
and the corresponding equation for $\chi_I$. 

We draw $\Delta (p/T^4)$ for each fixed $\mu_q/T$ in Fig.~\ref{fig:pres} 
(right) and find that the difference from $p|_{\mu_q=0}$ is very small 
in the interesting regime for heavy-ion collisions, 
$\mu_q/T \approx 0.1$ (RHIC) and $\mu_q/T \approx 0.5$ (SPS), 
in comparison with the value at $\mu_q=0$, e.g. the SB value for 2 flavor QCD 
at $\mu_q=0$: $p^{SB}/T^4 \simeq 4.06$. The effect of a non-zero quark density 
on the pressure at $\mu_q/T=0.1$ is only $1\%$. 
Comparing with previous result up to $O(\mu_q^4)$ \cite{us03}, the effect 
from $O(\mu_q^6)$ seems to be small in the range we investigated. 
Also, the result is qualitatively consistent with that 
of Ref.~\refcite{FKS} obtained by the reweighting method.

\begin{figure}[t]
\centerline{\epsfysize=1.6in\epsfbox{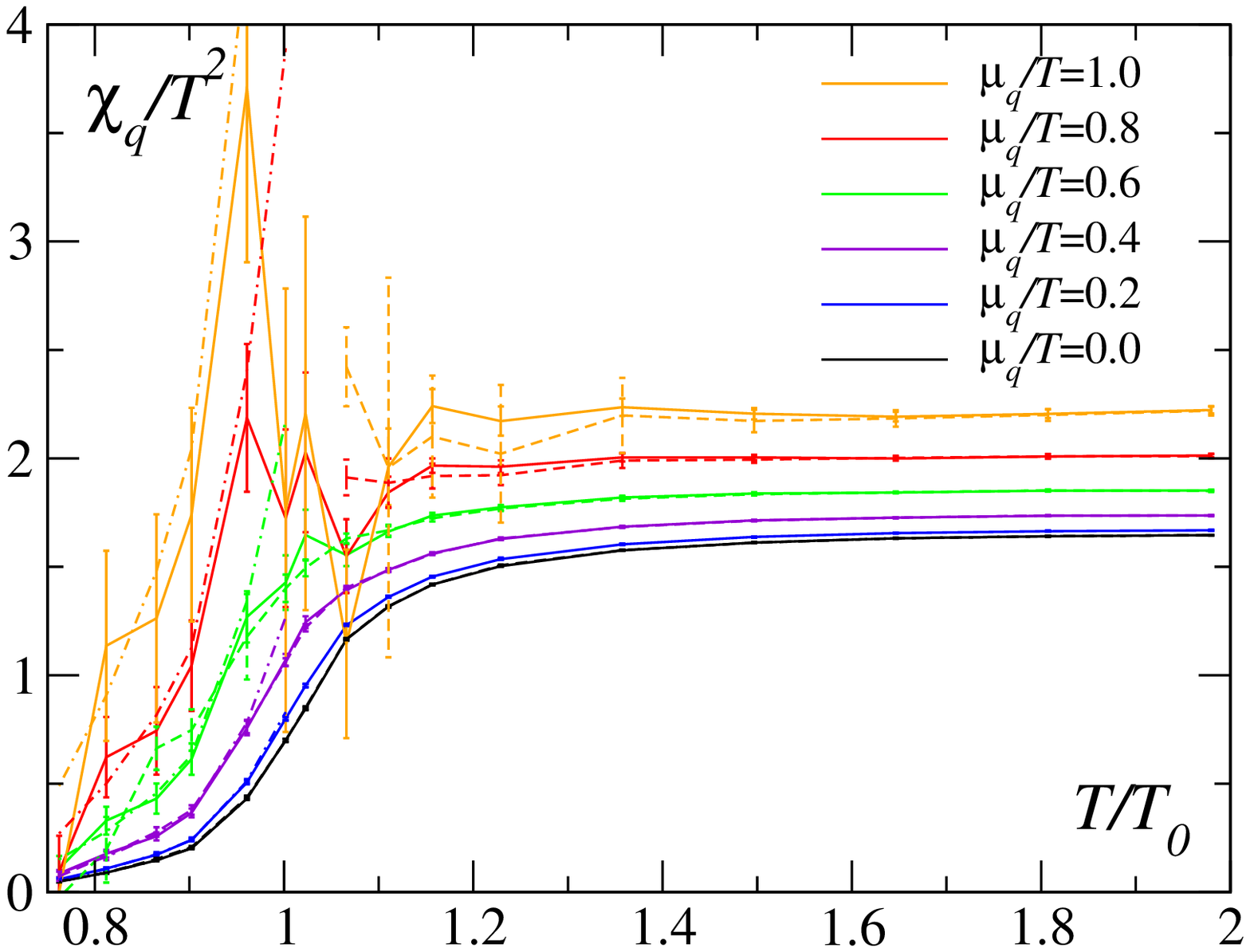}
            \epsfysize=1.6in\epsfbox{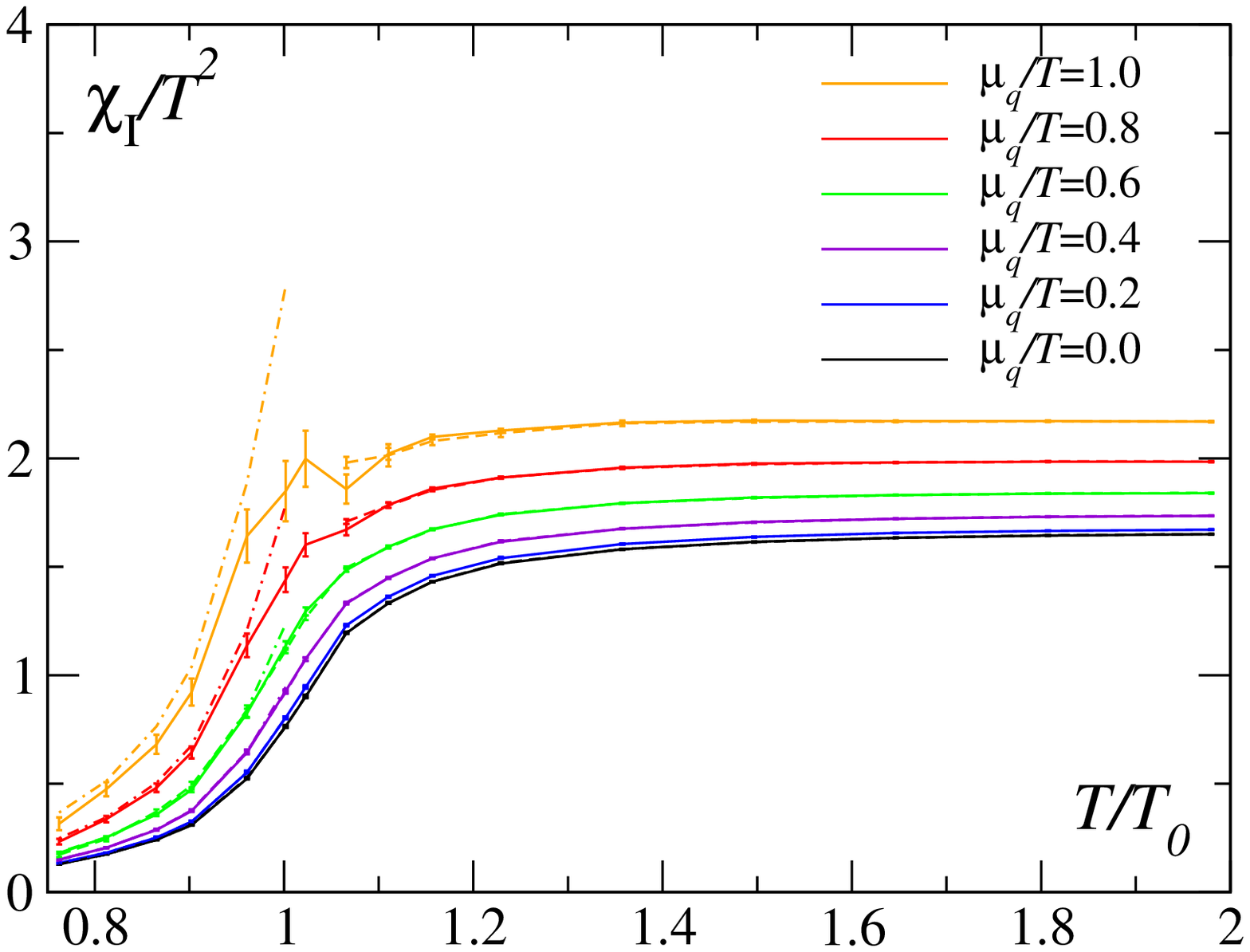}}   
\caption{
Quark number (left) and isovector (right) susceptibilities as 
a function of $T$ for each fixed $\mu_q/T$. 
$T_0$ is $T_c$ at $\mu_q=0$.
\label{fig:nsus}}
\end{figure}

The data connected by solid lines in Fig.~\ref{fig:nsus} 
are the results for $\chi_q$ (left) 
and $\chi_I$ (right) obtained by this method up to $O(\mu_q^4)$. 
We also calculated the susceptibilities by another two methods. 
Dot-dashed lines are from the hadron resonance gas model, using $F, F^I$ 
and $G^I$ we obtained. This turned out to be a good approximation 
for $T/T_c \simlt 0.96$. 
Dashed lines are calculated by the reweighting method using an 
approximation discussed in Ref.~\refcite{us02}. 
Here the truncation error is $O(\mu_q^6)$ but 
the effect from higher order terms of $\mu_q$ is 
partially included. The sign problem in the reweighting method 
is serious when the complex phase fluctuations of the quark 
determinant are large at large $\mu_q/T$. 
Hence we omitted data at which the standard deviation of the complex phase 
is larger than $\pi/2$. The difference among the three results is caused by 
the approximation in the higher order terms of $\mu_q/T$. 

Since the statistical error of $c_6$ is still large near $T_c$, the peak of 
$\chi_q$ is not clear. However, as seen in Fig.~\ref{fig:c2c4c6}, 
$c_6$ changes its sign at $T_c$. 
This means the peak position of $\chi_q$ moves left, 
which is corresponding to the change of $T_c$ as a function of $\mu_q$. 
$T_c(\mu_q/T=1)/T_c(\mu_q/T=0)$ in Ref.~\refcite{us02} is about $0.93$. 
Moreover, the strong enhancement of $\chi_q$ with increasing $\mu_q$ 
near $T_c$, obtained in Ref.~\refcite{us03}, is observed also 
in the analysis up to $O(\mu_q^4)$. 
This suggests the presence of a critical endpoint in the $(T,\mu_q)$ plane.
On the other hand, $\chi_I$ in Fig.~\ref{fig:nsus} does not show any 
singular behavior. 
This is consistent with the sigma model prediction that only isosinglet 
degrees of freedom become massless at the critical endpoint \cite{HS}.

\begin{thebibliography}{99}

\bibitem{us02}
C.R.~Allton {\it et al.}, {\it Phys. Rev.} {\bf D66}, 074507 (2002).

\bibitem{us03}
C.R.~Allton {\it et al.}, {\it Phys. Rev.} {\bf D68}, 014507 (2003).

\bibitem{KRT1}
F.~Karsch, K.~Redlich and A.~Tawfik, 
{\it Eur. Phys. J.} {\bf C29}, 549 (2003).

\bibitem{KRT2}
F.~Karsch, K.~Redlich and A.~Tawfik, 
{\it Phys. Lett.} {\bf B571}, 67 (2003).

\bibitem{HKS} U.M.~Heller, F.~Karsch, and B.~Sturm, 
{\it Phys. Rev.} {\bf D60}, 114502 (1999).

\bibitem{KLP}
F.~Karsch, E.~Laermann and A.~Peikert, 
{\it Phys. Lett.} {\bf B478}, 447 (2000).

\bibitem{FKS}
F.~Csikor {\it et al.} {\it JHEP} {\bf 0405}, 046 (2004). 

\bibitem{HS}
Y.~Hatta and M.A.~Stephanov, {\it Phys. Rev. Lett.} {\bf 91}, 102003 (2003). 





\end{thebibliography}
\end{document}